\documentclass{article}
\usepackage{iclr2027_conference,times}

\usepackage{amsmath,amsfonts,bm}

\def\eqref#1{equation~\ref{#1}}

\def\1{\bm{1}}

\DeclareMathAlphabet{\mathsfit}{\encodingdefault}{\sfdefault}{m}{sl}
\SetMathAlphabet{\mathsfit}{bold}{\encodingdefault}{\sfdefault}{bx}{n}

\usepackage{hyperref}
\usepackage{url}
\usepackage{booktabs}
\usepackage{graphicx}
\usepackage{multirow}
\usepackage{amsmath}
\usepackage{amssymb}
\usepackage{microtype}
\usepackage{xcolor}
\usepackage{enumitem}
\usepackage{adjustbox}
\usepackage{makecell}

\newcommand{\dev}[2]{%
_{\textcolor{red}{\scalebox{0.60}{#1}}}%
^{\textcolor{green!50!black}{\scalebox{0.60}{#2}}}%
}

\title{Degrees of Freedom in Multi-Agent Coordination: Between Centralized and Decentralized Regimes}

\author{%
{\bfseries Yiqing PENG$^{1,\dagger}$, Hehai Lin$^{1,\dagger}$, Zixuan Wang$^{1}$, Dongyi Zhang$^{1}$, Zhengjie Zhu$^{2}$,}\\[0.15em]
{\bfseries Yi Han$^{1}$, Yangyang Zheng$^{1}$, Hao Wang$^{4}$, Qian Li$^{3}$, Chengwei Qin$^{1,*}$}\\[0.5em]
{\normalfont $^{1}$The Hong Kong University of Science and Technology (Guangzhou)}\\
{\normalfont $^{2}$Wake Forest University, $^{3}$Shandong University}\\
{\normalfont $^{4}$University of Science and Technology of China}\\
}

\iclrfinalcopy
\begin{document}
\maketitle
\lhead{Preprint}
\begingroup
\renewcommand{\thefootnote}{\fnsymbol{footnote}}
\footnotetext[2]{Equal contribution.}
\footnotetext[1]{Corresponding author: \texttt{chengweiqin@hkust-gz.edu.cn}.}
\endgroup

\begin{abstract}
Large language model (LLM)-based multi-agent systems (MAS) are commonly organized around two coordination regimes. Centralized coordination provides explicit planning and control but can create a bottleneck at a single coordination locus, whereas decentralized coordination offers local autonomy but must achieve effective collaboration through distributed decisions and experience. These complementary trade-offs motivate a systematic study of the space between the two regimes. We introduce \emph{Degrees of Freedom} (DoF), a continuous configuration framework that quantifies the size of the decision space delegated to downstream agents. DoF jointly controls the granularity of delegated work and, within each granularity regime, the target probability that each task unit is assigned to the decentralized branch. Its endpoints recover the two pure coordination regimes. Across multiple task domains, DoF reveals a hierarchy of organizational effects. Delegated granularity and within-regime routing produce structured but bounded profile reallocations, whereas task domain and retained experience are associated with stronger differences between the centralized- and decentralized-branch profiles. The resulting performance landscape is task-dependent: different domains favor different DoF configurations, while atomic-target conditions are generally more consistent across runs. In-domain retained experience also brings the two branch profiles closer together. The same controlled sweep reveals a task-dependent coordination landscape: some domains exhibit clearly higher-performing configurations, whereas others show flatter responses, and no single configuration dominates across domains. In five of the six evaluated state--dataset settings, an intermediate DoF configuration achieves higher mean accuracy than both pure endpoints, demonstrating the value of exploring the coordination space between them. Together, these findings provide empirical guidance for studying and designing intermediate coordination regimes in LLM-based MAS.
\end{abstract}

\section{Introduction}

Large language model (LLM)-based multi-agent systems (MAS) extend agentic problem solving by decomposing complex requests into manageable units, assigning them to agents with different capabilities, and coordinating their interactions during execution \citep{guo2024llmmultiagents, huang2026amaadaptivememorymultiagent, yu2026correctedmemoryexecutionconsistency}. From the perspective of task allocation and decision authority, automatic MAS can be organized around two coordination regimes. Centralized coordination retains planning, assignment, review, and aggregation in an explicit controller, whereas decentralized coordination distributes execution and local interaction decisions among active agents. These regimes provide two distinct ways to organize the same underlying problem-solving process.

Centralized coordination benefits from a global view of the task and explicit control over agent selection and result integration, but concentrating these decisions at one locus can create planning, communication, and integration bottlenecks. Decentralized coordination supports local autonomy and distributed adaptation, but effective collaboration must emerge through successive local execution, splitting, and inter-agent transfer decisions \citep{cemri2025mast,mao2026delm,yang2025agentnet}. The resulting organization is therefore more dependent on interaction paths and accumulated agent state. The two regimes expose complementary advantages and structural costs rather than a universally preferable organization. Existing systems typically instantiate one complete architecture or optimize selected coordination mechanisms, leaving the space between centralized and decentralized coordination underexplored under matched experimental conditions.

\begin{figure*}[t]
    \centering
    \includegraphics[width=\textwidth]{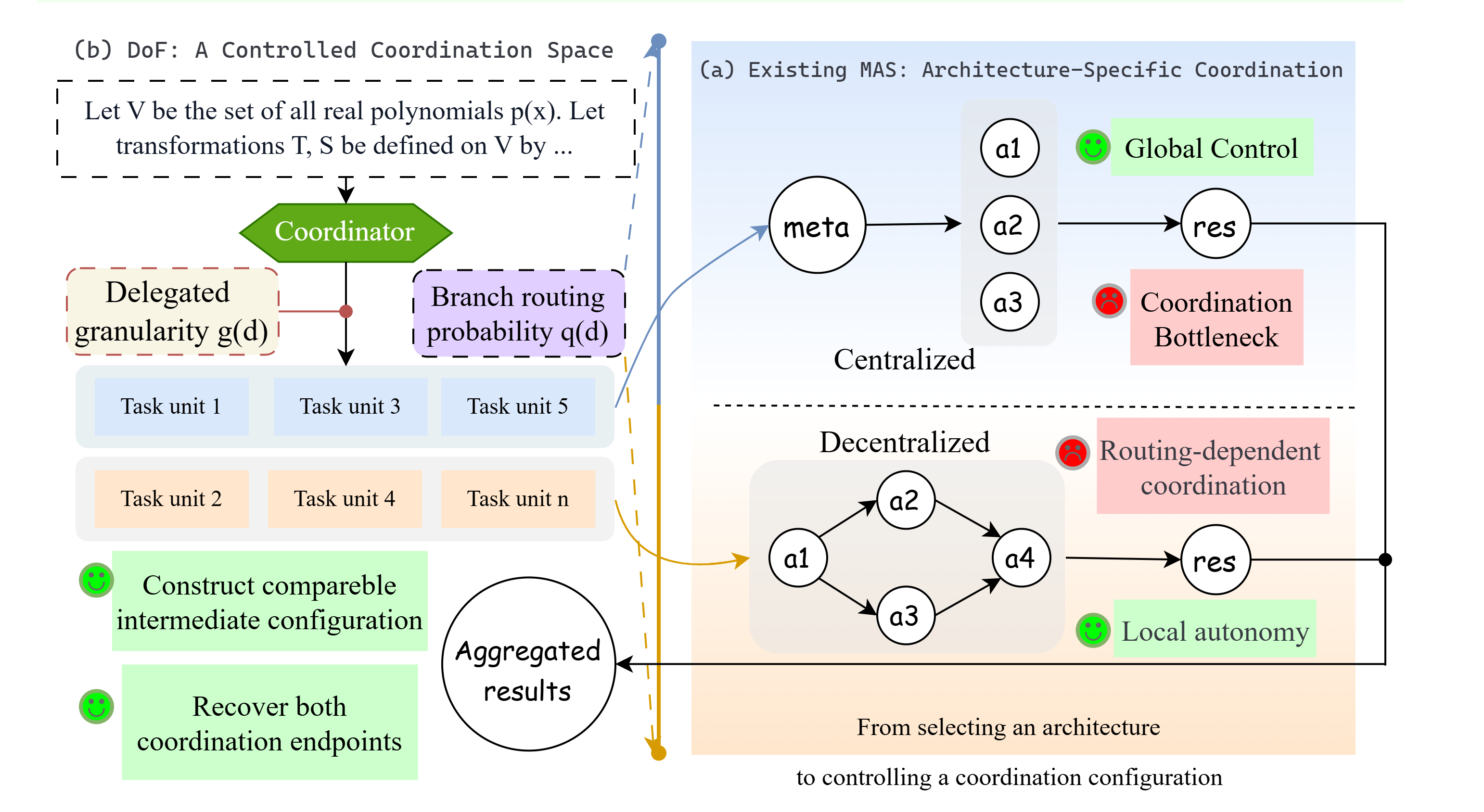}
     \caption{Illustration of the DoF framework. (a) Existing automatic MAS instantiate centralized or decentralized coordination through architecture-specific mechanisms, leaving intermediate configurations difficult to compare under matched conditions. (b) DoF moves from architecture selection to controlled coordination configuration by jointly specifying delegated granularity and per-unit branch-routing probability.}
    \label{fig:intro-dof-space}
\end{figure*}

We address this gap with the Degrees-of-Freedom (DoF) framework, illustrated in Figure~\ref{fig:intro-dof-space}, which interprets centralized and decentralized coordination as the two endpoints of a space of delegated decision authority. The configured degree of freedom $d\in[0,1]$ therefore indicates how much decision space is assigned to downstream agents: at $d=0$, the undecomposed request remains under centralized coordination, leaving no branch-level coordination autonomy to downstream agents, whereas at $d=1$, the original request is handled by decentralized coordination, exposing the full branch-internal decision space to the downstream agents. For $0<d<1$, the DoF Coordinator decomposes the request and controls downstream autonomy along two dimensions. First, delegated granularity determines the semantic scope of each unit: atomic-target units are narrower than subtask-target units and therefore give each downstream execution a smaller decision scope. Second, within either granularity regime, increasing the target probability of decentralized assignment gives more task units to the decentralized branch and enlarges the fraction of coordination decisions made downstream. Concretely, the atomic-target regime uses $0<d\leq0.5$ with $q(d)=2d$, whereas the subtask-target regime uses $0.5<d<1$ with $q(d)=2d-1$; thus, $q(d)$ increases with $d$ within each regime, while the granularity transition changes the scope of the delegated units. DoF consequently provides a single configuration index for allocating decision authority between the two endpoints without collapsing their distinct mechanisms into a third coordination regime.

We instantiate the centralized and decentralized branches with MetaAgent and AgentNet \citep{yang2025agentnet}, respectively, over the same five-agent roster and Qwen3-Next-80B \citep{qwen2025qwen3next80ba3binstruct}. We evaluate fifteen DoF configurations through repeated sweeps across five reasoning domains. This controlled design compares endpoint and intermediate configurations while keeping the underlying model, agent roster, and branch implementations fixed. The controlled sweeps reveal a progression from organizational adjustments to task outcomes. \emph{First}, delegated granularity and within-regime routing produce structured but bounded changes in worker-selection and node-activation profiles. \emph{Second}, task domain and retained experience are associated with stronger branch-specific differences: domain and execution branch establish distinct profiles, while in-domain retained experience brings the centralized- and decentralized-branch profiles closer together. The derived cooperation paths provide a descriptive view of the coordination events underlying these profiles. \emph{Finally}, the resulting performance landscape is task-dependent: intermediate configurations attain the highest observed mean in several settings, endpoint or near-endpoint configurations remain preferable in others, and atomic-target conditions are generally more consistent across runs. Together, these observations characterize coordination as a task- and state-conditioned design choice rather than a fixed preference for either endpoint. Our contributions are:
\begin{itemize}[leftmargin=*]
    \item \textbf{A controlled formulation of the coordination space.}
    We formulate centralized and decentralized coordination as endpoint regimes and identify delegated decision scope and processing locus as joint controls required to study the configurations between them.

    \item \textbf{A piecewise DoF coordination framework.}
    We introduce DoF as a common configuration index that combines two delegated-granularity regimes with per-unit branch-routing probability under a shared model, agent roster, and pair of branch implementations.

    \item \textbf{An empirical characterization of task-conditioned coordination.}
    Repeated DoF sweeps show that delegated granularity and within-regime routing produce structured but bounded changes in branch-specific profiles, while task domain and execution branch are associated with stronger profile differences and in-domain retained experience with closer cross-branch profiles. These organizational patterns accompany task-dependent accuracy peaks and greater run-to-run consistency for atomic-target conditions.
\end{itemize}

\section{Related Work}

\subsection{Automatic Multi-Agent Systems}

Centralized coordination regimes in automatic MAS place system construction and execution under an explicit controller or structured workflow. Existing frameworks program agent conversations, assign roles, decompose requests, select models or tools, and aggregate intermediate results \citep{wu2024autogen,li2023camel,hong2024metagpt,shen2023hugginggpt,chen2024agentverse}. Hierarchical approaches further combine high-level planning with specialized or dynamically constructed agents \citep{zhang2025agentorchestra,hou2025halo,wang2024megaagent}. More recent systems represent agents, dependencies, control flow, and message passing as executable graphs, or synthesize domain-specific nodes separately from topology orchestration \citep{liu2026masfactory,lin2026unifiedmas}. These designs enable explicit control over complex organizations, but retain coordination authority in a controller, hierarchy, or generated workflow.

Decentralized coordination regimes in automatic MAS instead distribute execution and interaction decisions among active agents. Dynamic interaction networks allow agents to select local actions and communication partners, so task allocation emerges from the evolving interaction graph rather than a single global plan \citep{liu2024dynamicllmpoweredagentnetwork,yang2025agentnet}. Shared-context approaches further distribute progress tracking, task claiming, and result integration across parallel agents \citep{mao2026delm}. Such systems reduce reliance on a single controller and support local adaptation, but their realized organization depends on sequences of local decisions, peer context, and retained agent state.

The two regimes therefore expose complementary limitations. Centralized systems can concentrate planning and integration at one coordination locus, whereas decentralized systems distribute control but make global organization dependent on local interaction and accumulated state. Selecting either regime alone does not reveal how much decision space should be delegated for a particular task. In this work, we introduce DoF as a unified coordination framework that represents centralized and decentralized coordination regimes as endpoints of a common configuration space and models intermediate systems through delegated granularity and per-unit branch routing. Our experiments show that intermediate configurations attain the highest observed accuracy in several task settings, while endpoint or near-endpoint configurations remain preferable in others, indicating that the appropriate coordination regime is task dependent.

\subsection{Empirical Analysis of Automatic Multi-Agent Systems}

Prior work analyzes automatic MAS through complementary diagnostic and intervention-based views. A useful organizing perspective treats MAS organization as a sequence of coupled decisions: what work is delegated, where it is routed, who performs it, how the resulting interaction unfolds, and how the organization adapts with experience. Task-decomposition methods determine the scope and granularity of delegated work, from recursively expanded sub-problems to structured reasoning units \citep{prasad-etal-2024-adapt,NEURIPS2023_271db992}, while compositional-routing work shows that decomposition granularity can condition downstream skill retrieval \citep{gao2026skillweaver}. Routing and workflow-construction methods then determine which agents, communication paths, or execution procedures receive each unit \citep{liu2024dynamicllmpoweredagentnetwork,yue2025masrouter}. This process is coupled with role assignment and agent specialization, which shape the capabilities and division of labor associated with each processing node \citep{hong2024metagpt,liu2024dynamicllmpoweredagentnetwork,yang2025agentnet}. Trajectory-level and behavioral studies further examine how roles are activated, how information propagates, and how interaction paths shape realized organization \citep{cemri2025mast,li-etal-2024-improving-multi,pmlr-v235-du24e,smit2024should,jwalapuram2026illusion}. Experience-based methods show that stored reflections, successful trajectories, agent-local memories, shared traces, selectively retained execution states, and transferable orchestration skills can modify later routing and interaction behavior \citep{shinn2023reflexion,zhao2024expel,ParkEtAl2023GenerativeAgents,gao2024inms,yang2025agentnet,rajib2026gatedmemory,lin2026skillmas}. Collectively, these analyses show that MAS behavior depends on topology, communication, decomposition, task assignment, branch-internal routing, model binding, and retained experience.

A complementary set of studies treats these organizational dimensions as intervention points, varying communication availability, local leadership, topology, collaboration protocol, orchestration structure, adaptive routing, delegation, or skill retrieval, and measuring changes in performance, cost, robustness, or consistency \citep{santos2023hybrid,sugawara2024lctt,pmlr-v235-zhuge24a,pmlr-v267-zhang25cu,zhang2025cut,yu2026adaptorch,chen2026organizationalscience,sander2026scaling,guan2026symphony,cui2026unoorchestra}. Existing work therefore provides important evidence about failures and adaptations, but typically studies one mechanism within a particular architecture or design family rather than a matched configuration space connecting centralized and decentralized coordination regimes. DoF advances this line of analysis by keeping the model, agent roster, and branch implementations fixed while sweeping fifteen endpoint and intermediate configurations. It uses branch-normalized allocation profiles, derived cooperation paths, repeated-run accuracy, and retained-state comparisons to characterize how organizational behavior and task outcomes vary across this controlled space, rather than introducing another endpoint-specific failure taxonomy.
\section{Degrees of Freedom Framework}
\label{sec:method}

\subsection{Problem Formulation and Overview}

Given an input task $x$ and a configured degree of freedom $d\in[0,1]$, the framework produces a final output $\hat{y}$. The endpoints retain the original request: $d=0$ sends $x$ directly to centralized coordination, whereas $d=1$ sends $x$ directly to decentralized coordination. For an intermediate configuration $0<d<1$, DoF specifies both the target granularity
\begin{equation}
 g(d)=
 \begin{cases}
 \mathrm{atomic}, & 0<d\leq 0.5,\\
 \mathrm{subtask}, & 0.5<d<1,
 \end{cases}
 \label{eq:granularity}
\end{equation}
and the target probability that a task unit is assigned to the decentralized branch
\begin{equation}
 q(d)=
 \begin{cases}
 2d, & 0<d\leq 0.5,\\
 2d-1, & 0.5<d<1.
 \end{cases}
 \label{eq:routing}
\end{equation}
Here, $g(d)$ determines the semantic scope of the delegated units, and $q(d)$ is the target probability that an eligible task unit is assigned to the decentralized branch. Thus, $d$ is a continuous configuration index. Granularity changes discretely between two granularity regimes, while the branch-assignment probability varies continuously within each regime. Appendix~\ref{app:routing-fidelity} compares these target probabilities with the realized decentralized task-unit fractions.

\begin{figure*}[t]
    \centering
    \includegraphics[width=\textwidth]{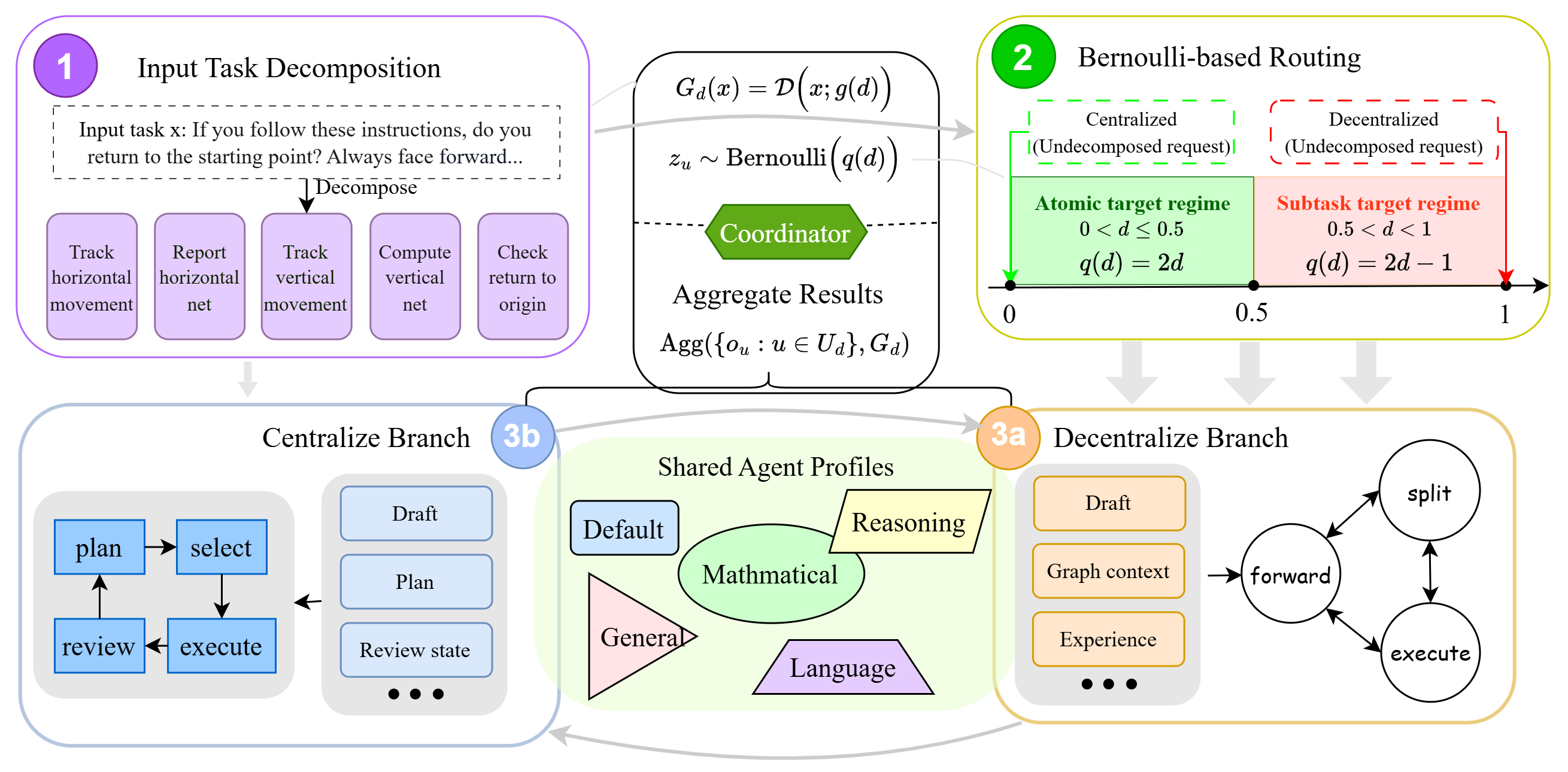}
    \caption{Overview of the DoF workflow. The Coordinator decomposes the input at the configured granularity, assigns each task unit to the centralized or decentralized branch through Bernoulli-based routing, and aggregates the completed unit outputs, linking the piecewise DoF schedule to the two branch implementations.}
    \label{fig:dof-framework}
\end{figure*}

For an intermediate configuration, the Coordinator decomposes the input into a dependency graph
\begin{equation}
 G_d(x)=(U_d,E_d)=\mathcal{D}\bigl(x;g(d)\bigr),
 \label{eq:decomposition}
\end{equation}
where $\mathcal{D}$ is the granularity-conditioned decomposition operation, $U_d$ is the set of routable task units, and $E_d\subseteq U_d\times U_d$ is the set of directed dependency edges. Once the predecessors of a unit $u\in U_d$ have completed, the Coordinator samples
\begin{equation}
 z_u\sim\mathrm{Bernoulli}\bigl(q(d)\bigr),
 \qquad
 z_u=
 \begin{cases}
 0, & \text{centralized branch},\\
 1, & \text{decentralized branch}.
 \end{cases}
 \label{eq:unit-routing}
\end{equation}
The execution context for $u$ is $h_u=\{(v,o_v):(v,u)\in E_d\}$, where $o_v$ denotes the completed output of a predecessor $v$. Let $\mathcal{B}_{\mathrm{C}}$ and $\mathcal{B}_{\mathrm{D}}$ denote the centralized and decentralized branch executors, respectively, and set $\mathcal{B}_0=\mathcal{B}_{\mathrm{C}}$ and $\mathcal{B}_1=\mathcal{B}_{\mathrm{D}}$. The output of unit $u$ is then
\begin{equation}
 o_u=\mathcal{B}_{z_u}(u,h_u).
 \label{eq:unit-execution}
\end{equation}
After all required units have completed, a dataset-specific aggregation function $\operatorname{Agg}$ produces
\begin{equation}
 \hat{y}(x;d)=
 \begin{cases}
 \mathcal{B}_{\mathrm{C}}(x,\varnothing), & d=0,\\
 \operatorname{Agg}\!\left(\{o_u:u\in U_d\},G_d\right), & 0<d<1,\\
 \mathcal{B}_{\mathrm{D}}(x,\varnothing), & d=1.
 \end{cases}
 \label{eq:dof-output}
\end{equation}
Equation~\ref{eq:dof-output} summarizes the complete input-to-output flow. For intermediate conditions, DoF configures decomposition and per-unit branch routing. The selected branch executes each dependency-ready unit, and the Coordinator aggregates the resulting outputs. Figure~\ref{fig:dof-framework} illustrates this process.

\subsection{DoF-Controlled Execution}

\paragraph{Coordinator.}
The Coordinator is the control interface between the DoF configuration and the two branch implementations. For intermediate conditions, it applies the configured granularity, tracks task units and dependencies, and releases each unit when its predecessors finish. It supplies predecessor outputs as context and routes each unit to a branch according to the configured probability, supporting sequential and parallel profiles. At the two endpoints, it bypasses intermediate decomposition and routing and sends the original request directly to the corresponding branch. When either branch completes a unit, the Coordinator records its output and exposes it to downstream units whose dependencies have become satisfied. After all required units finish, the dataset-specific aggregation procedure combines the unit outputs into the final answer.

\paragraph{Centralized branch}
We instantiate the centralized branch with MetaAgent, which receives a task unit and context and executes a bounded plan--select--execute--review loop. It constructs a compact plan, selects workers using their task-relevant capabilities and success summaries, and reviews each result and the remaining plan. Insufficient results trigger a redo or replan; otherwise, the accumulated results are synthesized into the unit output.

\paragraph{Decentralized branch}
We instantiate the decentralized branch with AgentNet \citep{yang2025agentnet}. AgentNet receives the assigned unit and its available predecessor context. At each internal decision point, the active agent selects one of three operations: \texttt{execute}, \texttt{split}, or \texttt{forward}. The router conditions this choice on the current request, the active agent's capabilities, neighboring-agent information, graph weights, and retrieved experience. \texttt{execute} produces a result locally, \texttt{split} creates further internal work for the decentralized branch, and \texttt{forward} transfers the request to another node. The Coordinator's decomposition determines the initial semantic scope of a delegated unit, whereas \texttt{split} is an AgentNet-internal operation that can further divide that unit during execution. The retained AgentNet state includes capability vectors, graph edge weights, router experience, and executor experience.

Thus, DoF controls how work is delegated and where each unit is processed, while MetaAgent and AgentNet retain their respective centralized and decentralized execution mechanisms.

\subsection{Trace-Derived Organizational Measurements}

The preceding subsections define how the framework produces an output. This subsection defines how its recorded coordination events are summarized. Let $c\in\mathcal{C}$ denote an evaluated condition, $r\in\mathcal{R}=\{\mathrm{C},\mathrm{D}\}$ denote the centralized or decentralized branch, and $a\in\mathcal{P}$ denote a profile in the shared profile set $\mathcal{P}$. For condition $c$, branch $r$, and profile $a$, let $n_{c,r}(a)$ be the corresponding event count. Centralized events count worker-profile selections, whereas decentralized events count node activations along decentralized traces. When branch $r$ contains at least one recorded event, its normalized branch-specific profile distribution is
\begin{equation}
 p_{c,r}(a)=\frac{n_{c,r}(a)}{\sum_{a'\in\mathcal{P}}n_{c,r}(a')}.
 \label{eq:allocation-profile}
\end{equation}
Accordingly, cross-branch comparisons concern branch-specific profile composition rather than equality of work volume or specialization quality.

For two profile distributions $P$ and $Q$ aligned on the common support $\mathcal{P}$, we use base-2 Jensen--Shannon divergence \citep{lin1991divergence}:
\begin{equation}
 \operatorname{JSD}(P,Q)=H\!\left(\frac{P+Q}{2}\right)-\frac{H(P)+H(Q)}{2},
 \label{eq:jsd}
\end{equation}
where $H(P)=-\sum_{a\in\mathcal{P}}P(a)\log_2 P(a)$ is the base-2 Shannon entropy. Therefore, $\operatorname{JSD}(P,Q)\in[0,1]$. Conditions are aligned by common task identifiers before paired distribution comparisons are computed. %
\section{Experimental Setup}
\label{sec:setup}

\paragraph{Model and branch implementations.}
All experiments use Qwen3-Next-80B \citep{qwen2025qwen3next80ba3binstruct} and the same five-agent roster. The centralized and decentralized branches are instantiated with MetaAgent and AgentNet, respectively. AgentNet implements the decentralized coordination regime described in prior work \citep{yang2025agentnet}. The model, agent roster, branch implementations, dataset-specific dependency mode, aggregation rule, and evaluation procedure remain fixed across DoF conditions. DoF changes decomposition granularity, the resulting task-unit schedule, and per-unit branch assignments. Detailed agent configurations and runtime settings are provided in Appendix~\ref{app:implementation}; the stable prompt contracts are summarized in Appendix~\ref{app:prompt-contracts}.

\paragraph{Benchmarks and retained states.}
The primary retained state is trained on a BBH collection and evaluated on BBH \citep{suzgun-etal-2023-challenging}, AIME \citep{huggingfaceh4_2025_aime2024,opencompass2025aime,zhang2026aime}, MMLU-Pro \citep{NEURIPS2024_ad236edc}, RBench-T \citep{pmlr-v267-guo25r}, and SMFR \citep{jwalapuram2026illusion} collections. We additionally construct a retained state trained on an MMLU-Pro collection and evaluate it on MMLU-Pro. Training and evaluation items are disjoint in every setting. Each retained state is loaded before an evaluation sweep and remains fixed throughout that sweep. 

\paragraph{DoF configurations and repeated sweeps.}
Each dataset is evaluated with three repeated DoF sweeps (MMLU-Pro-trained has two). Each sweep evaluates fifteen DoF values covering the centralized endpoint, the decentralized endpoint, the atomic-target granularity regime, and the subtask-target granularity regime. 
The fifteen evaluated DoF values are
\[
\{0,0.0001,0.1,0.2,0.25,0.4,0.5,0.5001,0.6,0.7,0.75,0.85,0.9,0.9999,1\}.
\]
The matched target-routing pairs used for granularity comparisons are
\[
(0.0001,0.5001),\ (0.1,0.6),\ (0.2,0.7),\ (0.25,0.75),\ (0.4,0.9),\ (0.5,0.9999).
\]
These pairs have matched target probabilities of task-unit assignment to the decentralized branch. The evaluation-allocation matrix is provided in Appendix~\ref{app:dataset-protocol}.

\section{Results}

We organize the results as a progression from coordination configuration to organizational conditions and task outcomes. We first measure the profile changes associated with delegated granularity and within-regime routing, then compare the stronger organizational differences associated with task domain and retained experience, and finally relate these profiles to task-dependent accuracy peaks and run-to-run consistency.

\begin{figure*}[t]
    \centering
    \includegraphics[width=\textwidth]{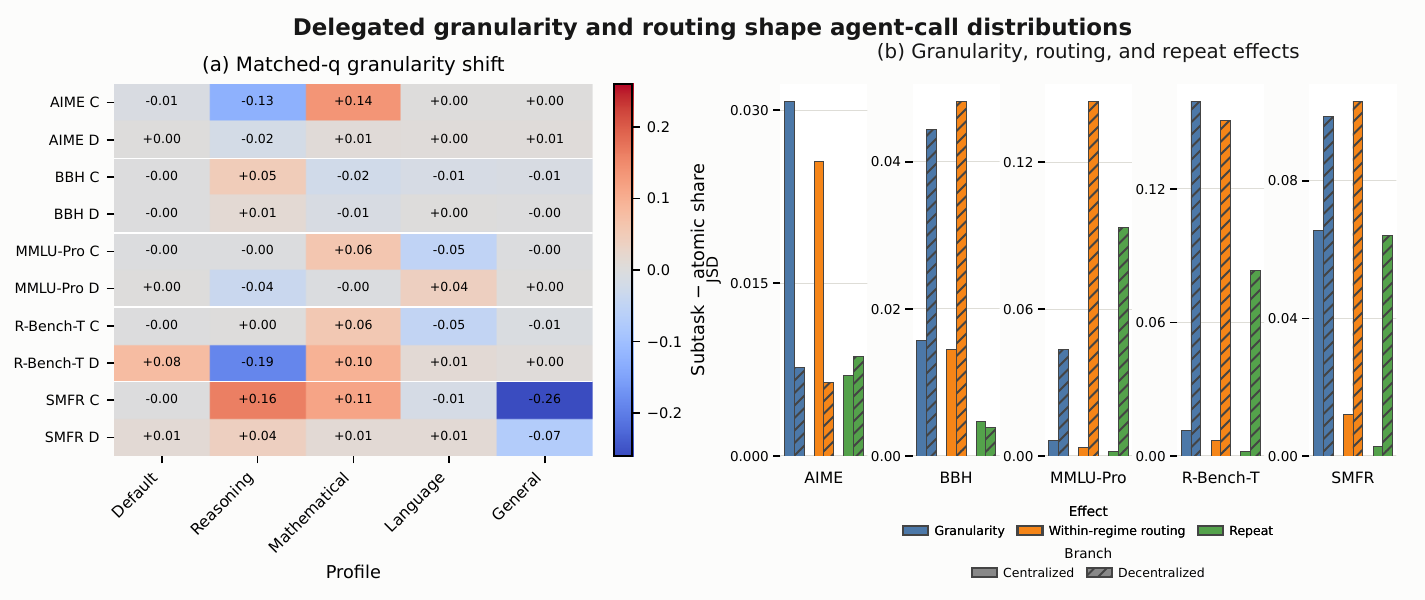}
    \caption{Delegated granularity and routing affect branch-specific profiles. Panel (a) reports the change in branch-specific profile composition from atomic-target to subtask-target delegation. Panel (b) compares matched-$q$ granularity differences, within-regime routing differences, and same-DoF repeated-run effects. Colors identify the effect type, while opacity distinguishes the centralized and decentralized branches.}
    \label{fig:granularity-routing}
\end{figure*}

\subsection{Delegated granularity and routing affect branch-specific profiles}

Figure~\ref{fig:granularity-routing} compares these two interventions using normalized profile distributions within each execution branch. The matched-$q$ comparison contrasts atomic-target and subtask-target conditions at the same target branch-assignment probability. Values are computed as the subtask-target share minus the atomic-target share. The within-regime comparison contrasts DoF values while holding the granularity regime fixed. These comparisons assess profile differences under the stated controls. 

\textbf{Changing delegated granularity produces structured but limited profile reallocation.} In Figure~\ref{fig:granularity-routing}(a), the subtask-target minus atomic-target differences are concentrated in selected profiles and dataset--branch pairs. The clearest signed changes occur in the centralized AIME, RBench-T, and SMFR rows, whereas BBH and MMLU-Pro remain closer across the matched pairs. Changing the semantic scope of a delegated unit does not erase the branch's existing profile preference. Instead, it changes how that preference is expressed for particular task domains. The effect is therefore consistent with granularity acting as a conditioning signal on an already learned branch-specific profile rather than creating a new organization independently of task and branch.

\textbf{Changing DoF within a fixed granularity regime has a comparable, also bounded effect.} Figure~\ref{fig:granularity-routing}(b) places within-regime routing JSD beside matched-$q$ granularity JSD and repeated-run JSD. The within-regime and granularity differences are of similar order for several dataset--branch pairs, indicating that changing the target branch-routing probability can affect a branch-specific profile even without crossing the atomic--subtask boundary. At the same time, all plotted JSD values are below $0.2$. Both interventions therefore produce modest reallocations within broadly similar distributions rather than transitions between qualitatively different branch-specific profiles. The repeated-run bars provide the reference scale: configuration changes are observable but remain bounded relative to the overall branch-specific profile. This result suggests that DoF tunes an existing coordination landscape rather than replacing one stable branch-specific profile with another. These bounded configuration effects motivate the next comparison: whether task domain and retained experience impose stronger organizational conditions on the branch-specific profiles.

\begin{figure*}[t]
    \centering
    \includegraphics[width=\textwidth]{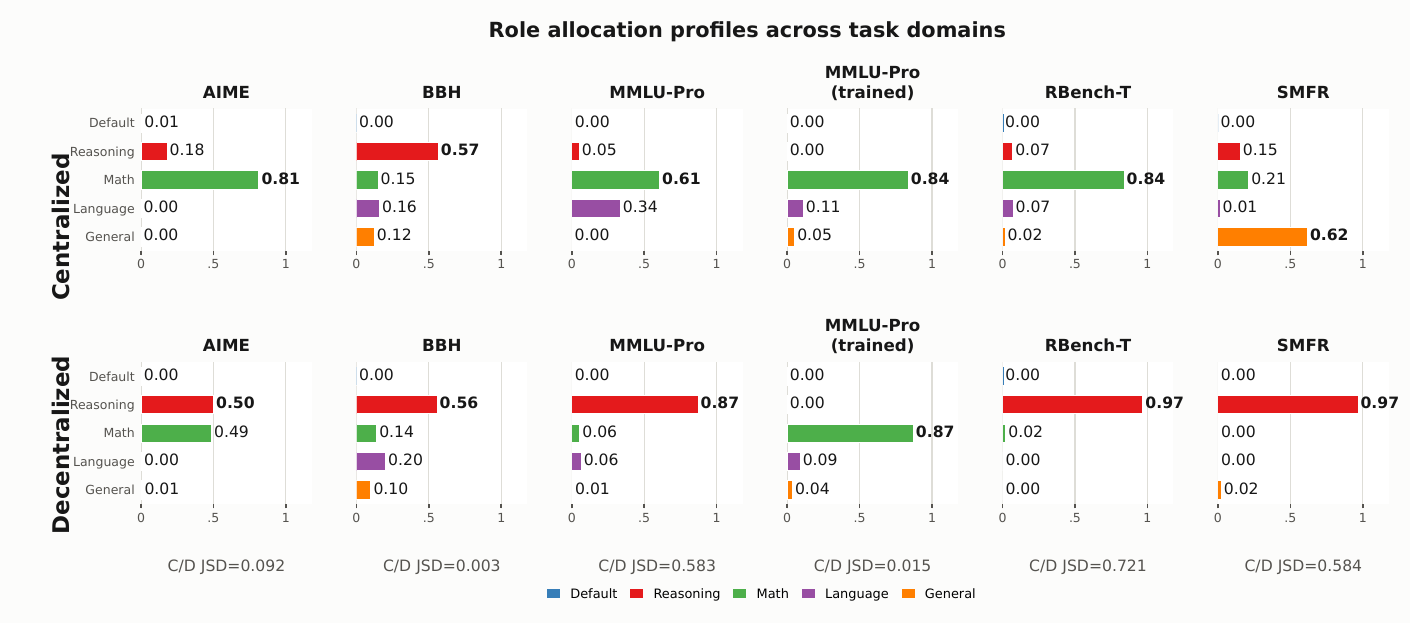}
    \caption{Branch-specific profiles across task domains and retained agent states. Bars show branch-normalized profile-call shares pooled over intermediate DoF conditions ($0<d<1$). The upper row reports centralized worker-profile selections, and the lower row reports decentralized node activations. The two MMLU-Pro columns distinguish the BBH-trained and MMLU-Pro-trained retained states. The branches are therefore compared by profile composition rather than by absolute workload volume.}
    \label{fig:role-allocation}
\end{figure*}

\subsection{Branch-specific profiles vary across task domains and retained states}

\textbf{Task domain and execution branch are associated with distinct profile compositions.} Figure~\ref{fig:role-allocation} shows branch-normalized profile shares pooled over intermediate DoF conditions. Mathematical profiles dominate centralized AIME and RBench-T, centralized SMFR emphasizes the general profile, and decentralized RBench-T and SMFR concentrate on reasoning. These differences are larger and more systematic than the modest within-regime shifts in Figure~\ref{fig:granularity-routing}. DoF adjusts the branch-specific profile, but task domain and execution branch provide the stronger organizational conditions under which those adjustments occur.

\textbf{Retained experience is associated with a reorganization of branch-specific profiles.} On the same MMLU-Pro evaluation set, the BBH-trained state gives a centralized--decentralized JSD of approximately $0.583$, whereas the MMLU-Pro-trained state reduces this value to approximately $0.015$. The in-domain retained state therefore brings the two branch-specific profiles substantially closer. This reduction is large relative to the within-regime DoF effects, which suggests that retained experience can reorganize the relationship between branches more strongly than a local change in target branch-routing probability. The derived cooperation paths provide a descriptive view consistent with this aggregate profile comparison, but do not establish a component-wise causal mechanism. Together with the domain differences above, this result indicates that task domain and retained experience impose stronger organizational conditions than the local granularity and routing adjustments examined in the first subsection.

\begin{table*}[t]
\centering
\scriptsize
\setlength{\tabcolsep}{3pt}
\renewcommand{\arraystretch}{1.15}

\begin{tabular}{l|cccccc}
\toprule
&
BBH &
AIME &
\makecell{MMLU-Pro\\(BBH-trained)} &
\makecell{MMLU-Pro\\(MMLU-Pro-\\trained)} &
R-Bench-T &
SMFR \\
\midrule

$d=0$
& $0.217\dev{-0.017}{+0.013}$
& $0.619\dev{-0.019}{+0.037}$
& $0.662\dev{-0.007}{+0.004}$
& $0.705\dev{-0.003}{+0.003}$
& $0.558\dev{-0.052}{+0.026}$
& $0.243\dev{-0.014}{+0.007}$ \\

$d=0.0001$
& $0.750\dev{-0.010}{+0.010}$
& $0.600\dev{-0.022}{+0.022}$
& $0.824\dev{-0.009}{+0.009}$
& $0.839\dev{-0.018}{+0.018}$
& $0.633\dev{-0.026}{+0.022}$
& $\mathbf{0.411}\dev{-0.036}{+0.036}$ \\

$d=0.1$
& $0.807\dev{-0.007}{+0.003}$
& $0.611\dev{-0.033}{+0.022}$
& $0.832\dev{-0.010}{+0.013}$
& $0.827\dev{-0.024}{+0.024}$
& $0.690\dev{-0.012}{+0.006}$
& $0.378\dev{-0.056}{+0.069}$ \\

$d=0.2$
& $0.813\dev{-0.003}{+0.007}$
& $0.607\dev{-0.030}{+0.015}$
& $0.835\dev{-0.013}{+0.010}$
& $0.842\dev{-0.015}{+0.015}$
& $0.704\dev{-0.014}{+0.022}$
& $0.396\dev{-0.010}{+0.010}$ \\

$d=0.25$
& $0.815\dev{-0.005}{+0.005}$
& $0.594\dev{-0.028}{+0.028}$
& $0.838\dev{-0.010}{+0.007}$
& $\mathbf{0.848}\dev{-0.009}{+0.009}$
& $0.692\dev{-0.014}{+0.010}$
& $0.375\dev{-0.052}{+0.042}$ \\

$d=0.4$
& $0.810\dev{-0.040}{+0.060}$
& $0.596\dev{-0.019}{+0.015}$
& $0.842\dev{-0.003}{+0.003}$
& $0.839\dev{-0.006}{+0.006}$
& $0.696\dev{-0.006}{+0.006}$
& $0.399\dev{-0.014}{+0.017}$ \\

$d=0.5$
& $0.883\dev{-0.023}{+0.017}$
& $0.581\dev{-0.004}{+0.007}$
& $\mathbf{0.856}\dev{-0.004}{+0.007}$
& $0.839\dev{-0.006}{+0.006}$
& $0.698\dev{-0.026}{+0.022}$
& $0.344\dev{-0.010}{+0.010}$ \\

$d=0.5001$
& $0.747\dev{-0.037}{+0.023}$
& $0.659\dev{-0.037}{+0.052}$
& $0.841\dev{-0.007}{+0.010}$
& $\mathbf{0.848}\dev{-0.003}{+0.003}$
& $\mathbf{0.730}\dev{-0.046}{+0.050}$
& $0.365\dev{-0.031}{+0.021}$ \\

$d=0.6$
& $0.787\dev{-0.027}{+0.023}$
& $\mathbf{0.667}\dev{-0.011}{+0.011}$
& $0.818\dev{-0.021}{+0.021}$
& $0.830\dev{-0.003}{+0.003}$
& $0.704\dev{-0.020}{+0.022}$
& $0.354\dev{-0.052}{+0.031}$ \\

$d=0.7$
& $0.827\dev{-0.027}{+0.053}$
& $0.604\dev{-0.026}{+0.019}$
& $0.826\dev{-0.016}{+0.013}$
& $0.833\dev{+0.000}{+0.000}$
& $0.679\dev{-0.030}{+0.024}$
& $0.396\dev{-0.010}{+0.021}$ \\

$d=0.75$
& $0.840\dev{-0.030}{+0.040}$
& $0.633\dev{-0.011}{+0.011}$
& $0.827\dev{-0.024}{+0.030}$
& $0.839\dev{-0.012}{+0.012}$
& $0.710\dev{-0.014}{+0.022}$
& $0.385\dev{-0.042}{+0.021}$ \\

$d=0.85$
& $0.867\dev{-0.017}{+0.023}$
& $0.619\dev{-0.019}{+0.015}$
& $0.841\dev{-0.007}{+0.010}$
& $0.824\dev{-0.021}{+0.021}$
& $0.698\dev{-0.002}{+0.004}$
& $0.382\dev{-0.049}{+0.045}$ \\

$d=0.9$
& $0.873\dev{-0.023}{+0.017}$
& $0.563\dev{-0.041}{+0.037}$
& $0.830\dev{-0.015}{+0.015}$
& $0.830\dev{-0.003}{+0.003}$
& $0.708\dev{-0.024}{+0.018}$
& $0.389\dev{-0.024}{+0.028}$ \\

$d=0.9999$
& $0.887\dev{-0.017}{+0.023}$
& $0.581\dev{-0.026}{+0.041}$
& $0.839\dev{-0.012}{+0.006}$
& $0.586\dev{-0.033}{+0.033}$
& $0.683\dev{-0.034}{+0.020}$
& $0.295\dev{-0.201}{+0.122}$ \\

$d=1$
& $\mathbf{0.953}\dev{-0.003}{+0.007}$
& $0.481\dev{-0.026}{+0.041}$
& $0.841\dev{-0.020}{+0.016}$
& $0.643\dev{-0.179}{+0.179}$
& $0.554\dev{-0.006}{+0.006}$
& $0.042\dev{-0.010}{+0.010}$ \\

\midrule
$\mathrm{Var}_{\mathrm{subtask}}/
 \mathrm{Var}_{\mathrm{atomic}}$
& 1.108 & 1.854 & 3.183 & 1.174 & 2.432 & 4.012 \\

\bottomrule
\end{tabular}

\caption{Accuracy across the canonical 15 DoF conditions.
Cell deviations are observed run ranges around the mean, not
confidence intervals; red is mean-to-min and green is max-to-mean.}
\label{tab:accuracy-dof}
\end{table*}

The preceding comparisons establish the organizational side of the coordination landscape: local configuration changes are bounded, whereas task domain and retained experience are associated with stronger profile differences. We next examine how this task- and state-conditioned organization is reflected in accuracy peaks and run-to-run consistency.

\subsection{Coordination configurations affect accuracy consistency and task performance}

\textbf{The performance consequence of coordination is task dependent.} Across the 15 DoF values, no configuration dominates. BBH peaks at the decentralized endpoint, while AIME, both MMLU-Pro settings, RBench-T, and SMFR attain their highest observed mean accuracy between the endpoints. The MMLU-Pro-trained setting has two tied intermediate peaks. These results favor intermediate configurations in five of six settings, but do not imply that mixing branches causes the gains. The BBH-trained MMLU-Pro peak uses only the decentralized branch, whereas the SMFR peak is nearly centralized. Their advantage over undecomposed endpoints is consistent with a contribution from task decomposition, although changes in execution and aggregation prevent us from isolating that contribution. Together with the distinct branch-specific profiles across tasks and retained states, these peaks support task-conditioned configuration rather than a universally best DoF.

\textbf{Atomic-target conditions are generally more consistent across repeated runs than subtask-target conditions.} For the datasets with valid complete-run variance estimates, the subtask-to-atomic variance ratios  exceed one in every reported comparison. This pattern suggests that broader subtask delegation exposes more internal choices to the receiving branch, making small generation differences more likely to propagate into different outcomes. It also separates performance level from performance stability. An intermediate configuration may have a high mean accuracy without being the most reproducible configuration across runs.

\section{Conclusion}

In this work, we introduce Degrees of Freedom (DoF), a controlled framework that connects centralized and decentralized LLM-based multi-agent systems through the size of decision space delegated to downstream agents. Experiments establish a coherent sequence. Configuration changes modestly affect branch-specific profiles. Task domain, execution branch, and retained state provide the stronger organizational conditions. Execution traces show how these profiles unfold as coordination events. The resulting organization is associated with both task performance and stability across runs. Together, these findings provide a new aspect for creating, comparing and evaluating LLM-based MAS.
\bibliography{refs}
\bibliographystyle{iclr2027_conference}

\appendix
\section{Supplementary Roadmap}
\label{app:overview}

The supplementary material is organized as follows. Appendix~\ref{app:implementation} reports the implementation settings and the stable prompt contracts used by the Coordinator, MetaAgent, and AgentNet. Appendix~\ref{app:dataset-protocol} documents the datasets, retained states, repeated-sweep allocation, evaluated DoF conditions, and error-handling protocol. Appendix~\ref{app:mmlu-quota} gives the subject-level MMLU-Pro sampling quotas. Appendix~\ref{app:supplemental-results} evaluates how closely realized branch assignments follow the target DoF routing probabilities. Appendix~\ref{app:analysis-artifacts} inventories the derived analysis artifacts and states the limits of the currently available trace archive.
\section{Implementation Details and Prompt Contracts}
\label{app:implementation}

\subsection{Model and agent roster}

All experiments use Qwen3-Next-80B. The roster contains five agent profiles with IDs 0--4: default, reasoning, mathematical, language, and general. All profiles use the same model, AgentNet router interface, executor interface, memory mechanisms, and retrieval procedures. They differ only in their initialized capability vectors.

Each capability vector contains seven dimensions: reasoning, mathematical, language, knowledge, sequence, spatial, and inference. The default profile assigns 0.6 to every dimension. The reasoning, mathematical, and language profiles assign 1.0 to the corresponding dimension and retain 0.6 on the other dimensions. The general profile assigns 0.7 to knowledge, sequence, spatial, and inference and 0.6 to reasoning, mathematical, and language. These vectors provide capability information to branch-internal agent selection.

\subsection{Agent graph and retained state}

The AgentNet branch is initialized with a complete directed graph over the five agents. Each directed edge starts with weight 1.0. During retained-state training, the state can include capability values, task-specific success statistics, edge weights, AgentNet router experience, and executor experience. Capability values and success statistics inform action selection. Edge weights and router experience inform AgentNet's branch-internal routing decisions, including the selection among \texttt{execute}, \texttt{split}, and \texttt{forward}, while executor experience supports task execution.

Edge weights are updated from forwarding outcomes and execution time. Edges whose weights fall below the active threshold are removed from subsequent neighbor information. The primary evaluation sweeps load a retained state before evaluation. The loaded state remains fixed during each sweep.

\subsection{Execution and model-call settings}

The AgentNet configuration allows up to three forward hops and five executions per task. These limits bound internal forwarding and execution within the decentralized branch. The MetaAgent configuration allows up to twelve planning rounds and one redo for a step. These limits bound planning and revision within the centralized branch. The AgentNet router and executor retrieve three experiences by default, and the executor memory limit is 40.

The Coordinator and MetaAgent share an LLM client configured with a maximum of 2,048 output tokens and temperature 0.3. AgentNet calls the same underlying model with a maximum of 2,048 output tokens, temperature 0, and top-$p$ of 1.0. Thus, the model identity is shared across the framework, while the branch implementations retain their original decoding interfaces.

\subsection{Prompt contracts}
\label{app:prompt-contracts}

The prompts use structured interfaces so that decomposition, branch-internal decisions, and aggregation can be recorded consistently. The descriptions below report the stable instructions and output schemas. Dynamic task text, retrieved experiences, agent states, and prior results are inserted at runtime.

\paragraph{Coordinator decomposition.}
The decomposition prompt receives the current task description, the requested granularity (subtask or atomic operation), and the dataset dependency mode. For a sequential profile, it states that later units may use earlier results; for a parallel profile, it requires independent units. The prompt explicitly prohibits solving the task, emitting derivations, or returning intermediate or final numeric answers. It requests only short, instruction-like unit descriptions in a JSON list with fields \texttt{id} and \texttt{description}. If parsing fails, the Coordinator retries once with a JSON-only reminder. A second failure falls back to one unit equal to the current task description.

\paragraph{Coordinator aggregation.}
For LLM-based synthesis, the aggregation prompt receives the original task, the ordered unit descriptions, their completed outputs, and any dataset-specific answer-format constraint. It instructs the model to assemble or extract the final answer only from the supplied results and prohibits new computation. Profiles using last-leaf aggregation return the final leaf result directly and do not invoke this prompt.

\paragraph{MetaAgent planning and selection.}
The planning prompt asks for the fewest necessary steps and prefers a single step unless the request contains separable reasoning stages. Its JSON response contains \texttt{needs\_decomposition}, \texttt{estimated\_difficulty}, and an ordered \texttt{plan}. For each plan step, the selection prompt supplies the task type, full plan, current progress, and the agents' task-relevant abilities, success summaries, and current loads. It returns \texttt{agent\_id} and \texttt{reason}; invalid selections fall back to capability-based selection.

\paragraph{MetaAgent execution and review.}
The selected worker first produces a scoped draft and then performs a separate verification and refinement pass. Intermediate units are instructed not to answer the original task, whereas the final unit receives the dataset-specific output constraint. The review prompt checks reasoning, correctness, completeness, and scope. It returns one of three JSON decisions: \texttt{completed} with the final answer, \texttt{continue} with a confirmed conclusion and revised remaining plan, or \texttt{redo\_current} with a revised step and failure feedback.

\paragraph{AgentNet routing and execution.}
The AgentNet router receives the major task, current unit, completed progress, active-agent state, neighboring-agent information, and retrieved experiences. Its structured response contains \texttt{DECISION}, \texttt{REASON}, \texttt{EXECUTABLE}, \texttt{DELEGATE}, \texttt{DESCRIPTION}, and \texttt{NEXT\_AGENT\_ID}. The decision is normalized to \texttt{execute}, \texttt{split}, or \texttt{forward}; invalid decisions default to local execution. The executor first generates a structured \texttt{RESULT} thought and then executes either the current request or an internally split unit under the dataset answer constraints. A separate completion check returns whether the request is complete and, if not, which agent should continue.

\paragraph{Disclosure scope.}
The appendix reports the reusable contracts rather than task-populated prompt dumps. Runtime prompts can contain benchmark questions, retained experiences, intermediate generations, and agent-state values; configuration files can also contain service endpoints and checkpoint locations. These values are excluded because they are not necessary to specify the method and require separate license, privacy, and credential review before artifact release.

\subsection{Dataset profiles}

Dataset profiles define the decomposition mode, final-answer format, answer extraction rule, and aggregation strategy. BBH and MMLU-Pro use parallel task-unit dependencies. AIME, RBench-T, and SMFR use sequential dependencies. Dataset-specific answer extraction and correctness evaluation are applied after the final output is produced.
\section{Dataset and Protocol Details}
\label{app:dataset-protocol}

\subsection{Evaluation collections and sweep allocation}

The primary BBH-trained retained state uses a disjoint BBH training collection of 634 items. It is evaluated on BBH, AIME, MMLU-Pro, RBench-T, and SMFR. The additional MMLU-Pro-trained retained state uses a disjoint subject-proportional training sample of 600 items and is evaluated on a separate MMLU-Pro sample. Table~\ref{tab:experiment-matrix} summarizes the evaluation allocation used by the paper-facing analyses.

\begin{table*}[t]
\centering
\small
\begin{adjustbox}{max width=\textwidth}
\begin{tabular}{lllrrl}
\toprule
Retained state & Evaluation collection & Evaluation subset & Items & Summary runs & Dependencies \\
\midrule
BBH-trained & BBH & Held-out BBH & 100 & 3 & Parallel \\
BBH-trained & AIME & AIME 2024--2026 snapshots & 90 & 3 & Sequential \\
BBH-trained & MMLU-Pro & Validation sample & 168 & 3 & Parallel \\
MMLU-Pro-trained & MMLU-Pro & Validation sample & 168 & 2 & Parallel \\
BBH-trained & R-Bench-T & English, text-only & 168 & 3 & Sequential \\
BBH-trained & SMFR & Two-investor subset & 96 & 3 & Sequential \\
\bottomrule
\end{tabular}
\end{adjustbox}
\caption{Training-state and evaluation allocation for the repeated DoF sweeps, in the canonical analysis order. The summary-run cohort is $3/3/3/2/3/3$. Each sweep targets the same fifteen DoF conditions; failed or missing conditions remain explicit.}
\label{tab:experiment-matrix}
\end{table*}

The subject quotas for the MMLU-Pro adaptation are listed in Appendix~\ref{app:mmlu-quota}. Training and evaluation items are disjoint within each retained-state setting.

\subsection{Run and error handling}

The planned valid-run cohort sizes are the summary-run counts in Table~\ref{tab:experiment-matrix}. A condition with an incomplete planned cohort is retained in the condition-level record and excluded from the variance summaries. Failed or null conditions retain their error state, and missing accuracy is not imputed. Accuracy means use every valid observed run for a condition. The variance comparison includes a DoF condition only when its complete planned cohort is valid.

\subsection{Trace instrumentation and available records}

The execution runner supports a full per-task trace schema containing the task, DoF configuration, decomposition log, unit dependencies, branch assignments, branch-internal events, unit outputs, aggregation log, final result, and correctness. The analysis checkout used for the present paper retains condition summaries and derived task-level cooperation paths, but not the raw per-task trace JSON files, original question text, generated answer text, dependency graphs, or branch communication content. Consequently, the reported quantitative analyses use only fields preserved in the summaries and path records and do not reconstruct unavailable answer-level traces.

The derived records preserve the dataset/state setting, run, task identifier, DoF value, correctness, condition accuracy, encoded cooperation path, and path length. Condition summaries additionally preserve the task count, average number of routed units, realized centralized assignment fraction, and any condition-level error. These fields support task alignment, repeated-run analysis, matched-routing comparisons, branch-profile summaries, routing-fidelity checks, and retained-state comparisons.
\section{MMLU-Pro Sampling Quotas}
\label{app:mmlu-quota}

The 600-item training and 168-item evaluation samples preserve the original fourteen-category proportions using largest-remainder allocation (Table~\ref{tab:mmlu-quota}). The two samples are disjoint.

\begin{table}[h]
\centering
\small
\begin{tabular}{lrr@{\qquad}lrr}
\toprule
Subject & Train & Eval. & Subject & Train & Eval. \\
\midrule
Math & 67 & 19 & Physics & 65 & 18 \\
Chemistry & 57 & 16 & Law & 55 & 15 \\
Engineering & 48 & 14 & Other & 46 & 13 \\
Economics & 42 & 12 & Health & 41 & 11 \\
Psychology & 40 & 11 & Business & 39 & 11 \\
Biology & 36 & 10 & Philosophy & 25 & 7 \\
Computer Science & 20 & 6 & History & 19 & 5 \\
\bottomrule
\end{tabular}
\caption{Subject-proportional MMLU-Pro sampling quotas.}
\label{tab:mmlu-quota}
\end{table}
\section{Supplemental Routing-Control Validation}
\label{app:supplemental-results}
\label{app:routing-fidelity}

\begin{figure*}[t]
    \centering
    \includegraphics[width=\textwidth]{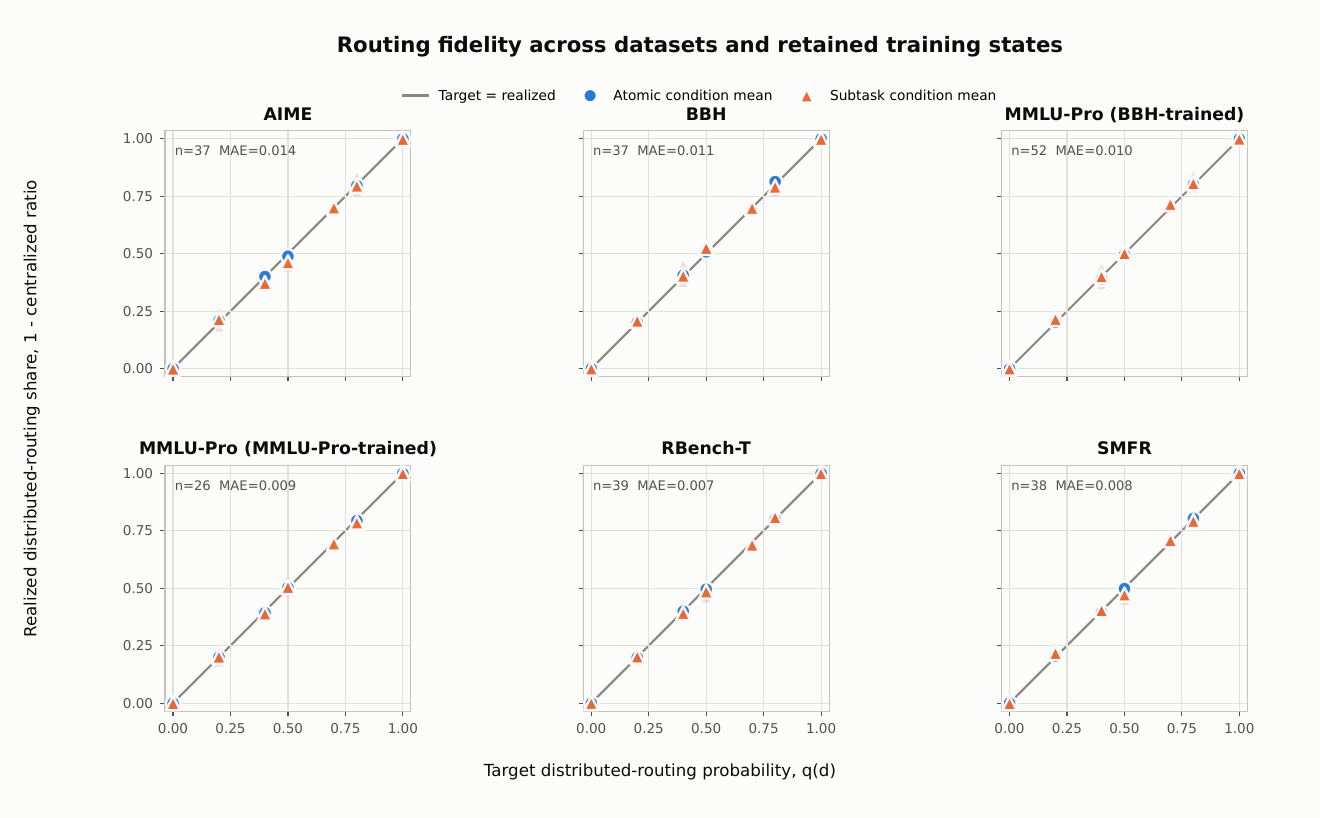}
    \caption{Target and realized decentralized task-unit assignment fractions across repeated DoF sweeps. Each panel represents one evaluation collection and retained-state setting. Small points show individual runs, large points show condition means, and the diagonal denotes exact agreement with the target $q(d)$. Marker shape distinguishes atomic-target and subtask-target conditions. Endpoint conditions are excluded because they bypass probabilistic per-unit routing.}
    \label{fig:routing-fidelity}
\end{figure*}

The DoF schedule specifies a target probability $q(d)$ that an eligible task unit is assigned to the decentralized branch. Because each task produces a finite number of routable units and assignments are sampled independently, the realized fraction can deviate from this target within a run. Figure~\ref{fig:routing-fidelity} compares the target probability with the realized decentralized fraction, computed as one minus the condition-level centralized assignment ratio, for every valid intermediate condition.

The comparison evaluates control fidelity rather than task accuracy. Across 216 valid intermediate run--condition points, the mean absolute deviation from the routing target is 0.0098. Deviations are expected when a condition contains a finite number of units, especially for coarser subtask decompositions that create fewer assignment draws. The machine-readable sidecar retains the complete point set and all 39 exclusions from the 255-condition planned matrix (34 endpoints, four failed intermediate conditions, and one missing intermediate condition), together with run identities and absolute-deviation summaries.
\section{Analysis Artifacts and Availability}
\label{app:analysis-artifacts}

The paper-facing analysis is derived from two surviving record types. Condition summaries provide accuracy, task count, average routed-unit count, realized centralized assignment fraction, and explicit condition errors. Task-level cooperation-path records provide dataset/state setting, run, task identifier, DoF value, correctness, encoded branch/profile path, and path length. The canonical task-condition collection contains 33,030 unique path records. Derived comparison tables contain 1,785 same-run DoF pairs and 328 matched-routing granularity pairs.

The validation report records 32,348 path records joined to valid condition summaries, 144 records associated with null or error summaries, 442 empty cooperation paths, and 96 records without matching summaries. These categories account for all 33,030 unique path records. Error and missing states remain explicit, and no missing accuracy is imputed. The analysis runner produces the canonical accuracy table, branch-specific profile tables, granularity and routing comparisons, repeated-run comparisons, routing-control validation, and machine-readable provenance sidecars.

The current analysis checkout does not contain the original per-task trace JSON files. It therefore cannot recover task text, decomposition responses, dependency graphs, model answers, review decisions, or communication content. The experiment runner defines and can write those fields, but any future release of raw traces must be inventoried separately and reviewed for benchmark redistribution terms, retained-memory content, service credentials, and user-specific paths. All claims in the present manuscript are restricted to the condition summaries and derived cooperation-path records available to the analysis pipeline.
\end{document}